\documentclass[copyright,creativecommons]{eptcs}
\providecommand{\event}{RTRTS 2010} % Name of the event you are submitting to
\usepackage{breakurl}             % Not needed if you use pdflatex only.

\usepackage{graphicx}
\usepackage{color}
\usepackage{alltt}
\usepackage{shortvrb}  % Ollie stuff
\usepackage{dsfont}    % Ollie stuff
\usepackage{amsmath}
\usepackage{amsfonts}
\usepackage{amssymb}
\usepackage{subfig}

\newcommand{\attone}{\mbox{$att_1$}}
\newcommand{\attn}{\mbox{$att_n$}}
\newcommand{\sone}{\mbox{$s_1$}}
\newcommand{\sn}{\mbox{$s_n$}}

\def\nnReals{\mathbb{R}_{\ge 0}}
\def\Naturals{\mathbb{N}}

\newcommand{\hide}[1]{}

\newcommand{\techrep}[1]{}    %% only in long paper
\newcommand{\shortpaper}[1]{#1}  %% only in short paper

\title{Extending the Real-Time Maude Semantics of Ptolemy to Hierarchical DE Models}
\author{Kyungmin Bae
\institute{Department of Computer Science \\ University of Illinois at Urbana-Champaign}
\and
Peter Csaba {\"O}lveczky
\institute{Department of Informatics \\  University of Oslo}
}
\def\titlerunning{Verifying Hierarchical Ptolemy II DE Models}
\def\authorrunning{K. Bae and P. C. {\"O}lveczky}
\begin{document}
\maketitle

\begin{abstract}
This paper extends our  Real-Time Maude formalization of the semantics
of \emph{flat} Ptolemy II discrete-event (DE) models to  \emph{hierarchical}
models, including \emph{modal} models. This is a challenging task that requires 
combining synchronous fixed-point computations with hierarchical structure. 
The synthesis of a Real-Time Maude verification model from a Ptolemy II DE model, 
and  the formal verification of the synthesized  model in Real-Time Maude,
have been integrated into Ptolemy II, enabling a model-engineering 
process that combines the convenience of Ptolemy II
DE modeling and simulation with formal verification in Real-Time Maude.

\end{abstract}

\section{Introduction}

One of the most promising approaches to
increase the use of formal methods 
 is to enrich the intuitive, often graphical, \emph{informal} modeling languages
preferred by practitioners with formal analysis capabilities by: (i) providing a formal semantics
to such informal  languages, (ii) automatically translating a model in the 
informal language into a formal model, and then (iii) verifying the formal model.  

For \emph{real-time} systems, we believe that
 \emph{real-time rewrite theories}~\cite{OlvMesTCS} 
should be a suitable formalism in which to define the semantics
of time-based modeling languages, for the following reasons:
\begin{itemize}
\item Real-time rewrite theories have a natural and ``sound'' model of timed behavior
that makes them suitable as a semantic framework,  
and avoids having to prove theorems like  ``all ill-timed behaviors can be rearranged
into equivalent well-timed behaviors,'' as might be needed when using, say,  a timed Petri net 
semantics (see, e.g.,~\cite{gyapay-varro-heckel-03} for one such example).
\item  The expressiveness and generality of real-time rewrite theories allow us to 
give a formal semantics to languages with advanced functions and data types, 
new communication models,
arbitrary and unbounded data structures,  program 
variables ranging over unbounded domains, and so on. 
\item The associated Real-Time Maude tool~\cite{journ-rtm} provides a range 
of formal analysis
capabilities, including simulation, reachability analysis, and linear temporal logic model
checking. Despite  the expressiveness of  real-time rewriting, timed-bounded 
LTL properties are often decidable under mild conditions~\cite{wrla06}. 
\end{itemize}

Real-time rewrite theories and Real-Time Maude
 have  been used to define the formal semantics of -- and to provide
a simulator and model checker  for -- some real-time modeling languages, including:
a timed extension of the Actor model~\cite{rtactors-in-rtmaude}, 
the Orc web services orchestration language~\cite{musab-orc07}, 
a language 
developed at DoCoMo laboratories for handset applications~\cite{musab-fase09},
a behavioral subset of the avionics standard AADL~\cite{fmoods10},
the visual model transformation language e-Motions~\cite{emotions-wrla10},
 real-time model transformations in MOMENT2~\cite{fase10},
and \emph{flat} Ptolemy II  discrete-event models~\cite{icfem09}.

Ptolemy II~\cite{Lee:03:Ptolemy} is a well established modeling and simulation tool
used in industry.  A major   reason for its popularity is 
Ptolemy II's  powerful yet intuitive graphical modeling language that allows
a user to build hierarchical models that combine different models of computations. 
 In this paper, we focus on discrete-event (DE) models, which are explicit
about the timing behavior of systems. 
 The Ptolemy II DE models have a  semantics rooted in the fixed-point
semantics of synchronous languages~\cite{LeeZheng:07:SRDECT}.

Like many  graphical modeling languages, Ptolemy II DE models lack at
present formal verification capabilities. Furthermore, it seems that 
Ptolemy II DE models fall outside  the class of languages which can be given an 
automaton-based  semantics, because:  
(i) certain constructs, such as noninterruptible timers and ramps,  require
the use of data structures (such as lists) of unbounded size; 
(ii) the variables used in, e.g., the transition systems in FSM actors
range over infinite domains such as the  integers; and
(iii) executing a synchronous step  requires  fixed-point computations. 

In a recent paper~\cite{icfem09}, we presented  a formal semantics for  
a significant subset of \emph{non-hierarchical}  (or \emph{flat}) Ptolemy II DE models.
We have used  Ptolemy II's code generation 
infrastructure to automatically synthesize a Real-Time Maude 
verification model from a Ptolemy II model, and have integrated Real-Time Maude 
verification into  Ptolemy II, so that
Ptolemy II models can be formally analyzed from within Ptolemy II. 
This integration of Ptolemy II and Real-Time Maude
 enables a model-engineer\-ing process that combines
the convenience of Ptolemy II modeling with formal verification in
 Real-Time Maude. 

In this paper, we extend and modify the Real-Time Maude semantics of Ptolemy II 
to the much more complex transparent \emph{hierarchical} 
 Ptolemy II DE models, including \emph{modal}
models. We  define useful generic temporal logic propositions for
such  models, so that a Ptolemy II user can easily define his/her 
temporal logic requirements, from within Ptolemy II, without understanding Real-Time Maude
or the formal representation of a Ptolemy II model. 
We illustrate such  formal verification on a hierarchical
fault-tolerant traffic light control system. 

Our work on formalizing Ptolemy II is the first attempt to define
a Real-Time Maude semantics for \emph{synchronous} real-time languages. 
Apart from the important result of endowing hierarchical Ptolemy II DE models
with formal verification capabilities, the main contribution of this work is
to show how Real-Time Maude can  define the formal semantics
of synchronous real-time languages with  fixed-point semantics
and  hierarchical structure. These techniques should  be useful for 
defining the semantics of other
hierarchical synchronous languages. For example, motivated by the
 complexity-reducing PALS (physically asynchronous, logically synchronous) 
architecture pattern~\cite{pals-rtss09,pals-formal}, which allows us to verify
a  synchronous real-time system design while ensuring that the properties
also hold for the system's distributed asynchronous implementation,
there is currently an interest in extending the avionics modeling standard AADL~\cite{aadl}
to synchronous behavioral AADL models.
Since  AADL models are  hierarchical,    techniques
in this paper could carry over to the definition of a Real-Time Maude semantics
of such a synchronous version of AADL, endowing such AADL models
with verification capabilities.

Our work is conducted in the context of the NAOMI project~\cite{DJS08NAOMI}, 
where Lockheed Martin Advanced Technology Laboratories (LM ATL), UC Berkeley, UIUC, 
and Vanderbilt University  
work together to develop a multi-modeling design methodology. 
A key part of this project is the systematic use of model transformations and code 
generation to maintain consistency across models.

Section~\ref{sec:prelim}  introduces Ptolemy~II and 
Real-Time Maude. Section~\ref{sec:flat-ptolemy-semantics} recalls 
the Real-Time Maude semantics of \emph{flat} Ptolemy II DE models
described in~\cite{icfem09}. Section~\ref{sec:hierarchical-ptolemy} 
describes the hierarchical features~of Ptolemy II.
Section~\ref{sec:hierarchical-semantics}  shows how our semantics for flat models
has been extended to  hierarchical DE models. Section~\ref{sec:temp-logic} 
presents some useful predefined atomic propositions, allowing users to easily specify their
desired system requirements. Section~\ref{sec:case-study} illustrates  Real-Time-Maude-based 
verification in  Ptolemy II  with a hierarchical model of a fault-tolerant traffic light system.
Finally, Section~\ref{sec:related} presents
related work and gives some concluding
remarks.
More details about the Real-Time Maude semantics of
 Ptolemy, as well as  additional verification case studies, are given
 in the longer technical report~\cite{pt-rtm-techrep}.

\MakeShortVerb{\@}    %%  this means that @text@ is the same as \verb+text+
\section{Preliminaries \shortpaper{on Ptolemy II and Real-Time Maude}}
\label{sec:prelim}
\techrep{This section presents some preliminaries on 
the selected discrete event subset of Ptolemy (Section~\ref{sec:ptolemy-intro}) rewriting logic and
Real-Time 
Maude (Section~\ref{sec:rtm})  and
the selected discrete event subset of Ptolemy (Section~\ref{sec:ptolemy-intro}).\hide{, and the NAOMI
 multi-modeling infrastructure (Section~\ref{sec:naomi-intro}). }}           %% includes the following three as subsections:
\subsection{Ptolemy II and its DE Model of Computation}
\label{sec:ptolemy-intro}
%% Peter: this is stavros' version (I guess), but shortened for the conf paper. 

The Ptolemy project\footnote{\url{http://ptolemy.eecs.berkeley.edu/}} studies
modeling, simulation, and design of concurrent, real-time, embedded systems.
The key underlying principle in the project is the use of well-defined {\em
models of computation} (MoCs) that govern the interaction between concurrent
components. A major problem area being addressed is the use of {\em
heterogeneous} mixtures of MoCs~\cite{Lee:03:Ptolemy}. A result of the project
is a software system called Ptolemy~II, implemented in Java. Ptolemy~II allows
a user to build hierarchical models that  combine different
MoCs, including state machines, data flow, and discrete-event models. Models
can be graphically  designed and simulated. In addition, Ptolemy~II's  
\emph{code generation} capabilities
allow  models to be translated into models in other languages or into
imperative code, e.g.,  in C and~Java.

A Ptolemy~II model is a hierarchical composition of {\em actors} with {\em
connections} between the actors' {\em input ports} and {\em output ports}. The
actors represent data manipulation units, whose execution is governed by a
special attribute  called the {\em director}. 
%%The connections represent communication channels. 
%% Since the behavior of a model is
%% encapsulated and is well-defined with the director that it uses and the
%% actor-connection diagram, 
An essential feature of Ptolemy~II is hierarchy: a Ptolemy~II model
 can itself be treated as an actor, 
 called  a {\em composite actor}. 
%% (Non-composite actors are called {\em
%% atomic actors}.) 
%% This helps hide internal details of parts of a model, and therefore is
%% crucial for managing model complexity, and for achieving modularity and
%% scalability.  
Ptolemy~II also supports \emph{modal models}, which are 
finite state machines where each state of the
machine can be refined into an internal model.

Ptolemy II \emph{discrete-event} (DE) actors consume and 
produce {\em events} at their input and output ports. An
 event is a pair $(v,t)$ where $v$ is a {\em value} 
 and $t$ is a {\em tag}, modeling the time at which the
event occurs. Ptolemy~II DE models use {\em super-dense} time, in which a
tag $t$ is a pair $(\tau,n) \in \nnReals \times
\Naturals$, where $\tau$ is the {\em timestamp} that indicates the model time
when this event occurs, and $n$ is the {\em microstep index}. 

The semantics of Ptolemy~II DE models~\cite{Lee99DE} combines a
synchronous-reactive fixed-point iteration with advancement of time governed
by an event queue~\cite{LeeZheng:07:SRDECT}. Events in that queue are ordered
by their tags. Operation proceeds by iterations, each time removing the 
event(s) with the smallest tag from the queue.  The removed events are fed to their designated
actors. After that, actors with events available are executed, which may
generate new events into the queue.
A difference between Ptolemy~II and standard DE simulators is that, at any
model time $(\tau,n)$, the semantics is defined as the {\em least fixed-point}
of a set of equations, similarly to a synchronous
model~\cite{EdwardsLee03}. This allows Ptolemy~II models to
have arbitrary {\em feedback loops}. Semantics of such models can always be
given although they
may result in {\em unknown} ({\em bottom}) values, in case the model 
contains {\em
causality cycles}. 
  Conceptually, the semantics can be captured by the following pseudo-code:

\begin{small}
\begin{verbatim}
Q := empty; // Initialize the event queue to be empty.
for each actor A do A.init();  // Initialize A; may generate new events in Q
while Q is not empty do
  E := set of all events in Q with the smallest tag;
  remove E from Q;
  initialize ports with values in E or "unknown";
  while port values changed do
    for each actor A receiving new values do
      A.fire();   // May increase knowledge about presence/absence of inputs at ports
  end while;      // Fixed-point reached for the current tag
  for each actor A that has been fired do
    A.postfire(); // Updates actor state; may generate new events in Q
end while;
\end{verbatim}
\end{small}

%% To complete the semantics, the semantics of each actor needs to be defined.
%% For example, the {\em TimedDelay} actor has a single input port, a single output
%%	port, and a parameter $d\in\nnReals$. For each event $(v,(\tau,n))$ that it
%%	receives at its input port, the TimedDelay actor produces an event $(v,(\tau',n'))$ at its
%%	output port, such that: if $d>0$ then $\tau'=\tau+d$ and $n'=0$, otherwise,
%%	$\tau'=\tau$ and $n'=n+1$. 

%% The semantics is implemented with the DE director. That director can be
%% associated with a model to assign DE semantics to it. Besides DE director,
%% Ptolemy~II provides an extensible set of directors implementing different
%% semantics. In this work, we are also interested in {\em modal models}, which
%% are models with FSM (finite state machine) controllers. In spirit of
%% hierarchical heterogeneous modeling, Ptolemy~II allows composition between DE
%% models and modal models. The modal models capture dynamic states of the
%% systems, whereas the DE models manage time advance and processing of events.

\paragraph{Example: A Simple Traffic Light System.}
\label{sec:pt-example}

\begin{figure}[htb] \center
\includegraphics[width=125mm]{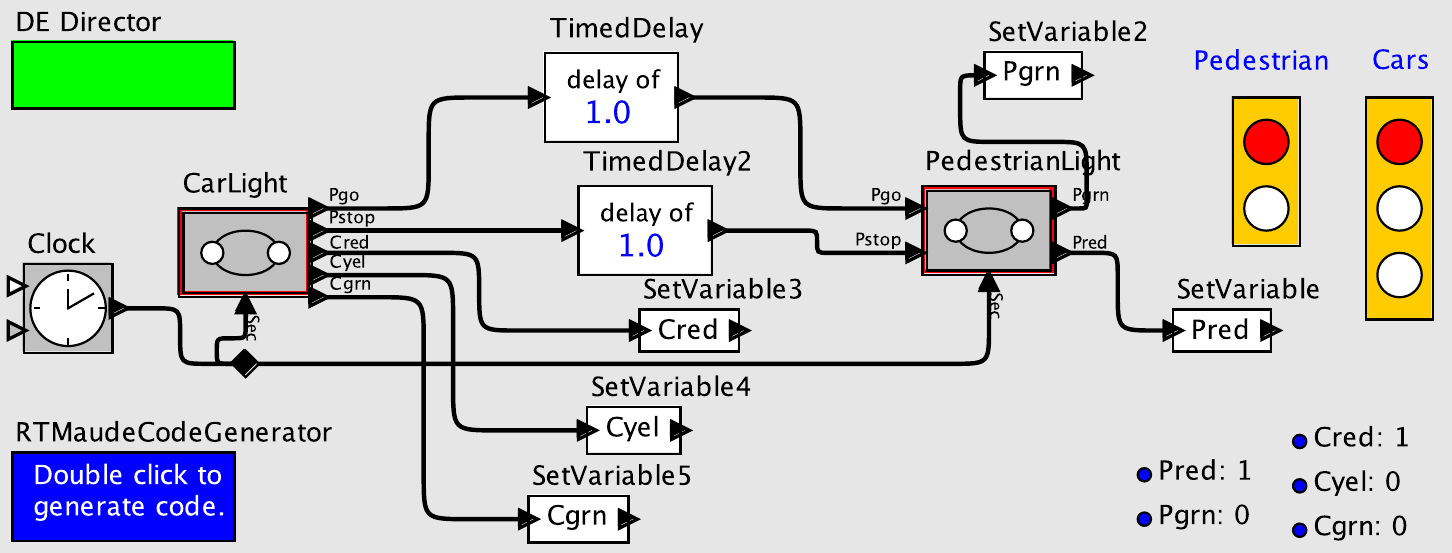}
\caption{\label{fig:tl1} A simple traffic light model in Ptolemy II.}
\end{figure}

Figure~\ref{fig:tl1} shows a  Ptolemy DE model of a simple 
traffic light system consisting of one car light and one pedestrian light
at a pedestrian crossing. Each  light is represented by a set
of  \emph{set variable} actors (@Pred@ and @Pgrn@ represent the pedestrian light,
and @Cred@, @Cyel@, and @Cgrn@ represent the car light). A light is 
 \emph{on} iff  the corresponding variable has the  value 1. 
The lights are controlled by  two \emph{finite state machine} (FSM)  
actors, \texttt{CarLight} and \texttt{PedestrianLight},
that send values to set the variables; in addition,  \texttt{CarLight}  sends
signals (that are \emph{delayed} by one time unit) to the \texttt{PedestrianLight} 
actor through its @Pgo@ and @Pstop@ output ports.

\begin{figure}[htb]
  \center
  \subfloat[][\texttt{PedestrianLight}]{\label{fig:tl2}\includegraphics[scale=0.7]{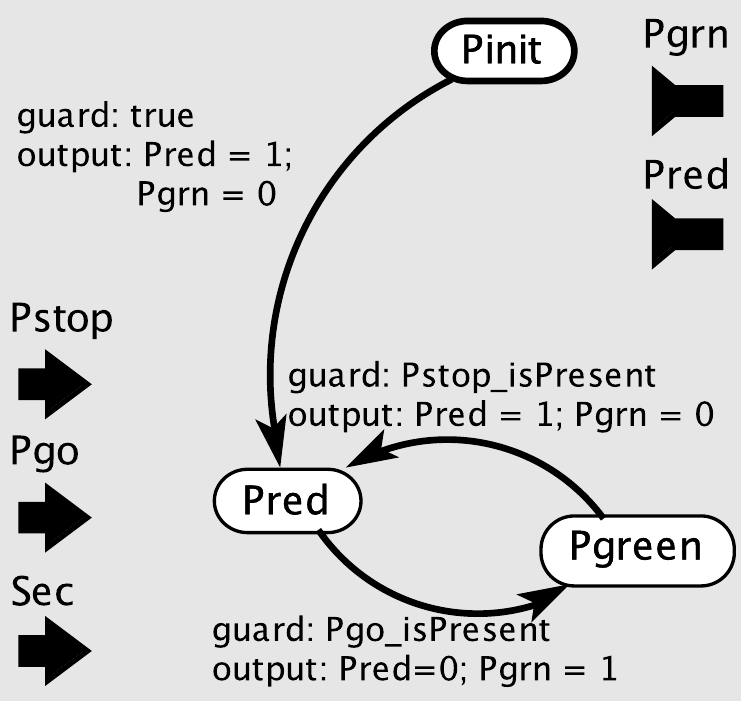}}
  \quad
  \subfloat[][\texttt{CarLight}]{\label{fig:tl3}\includegraphics[scale=0.68]{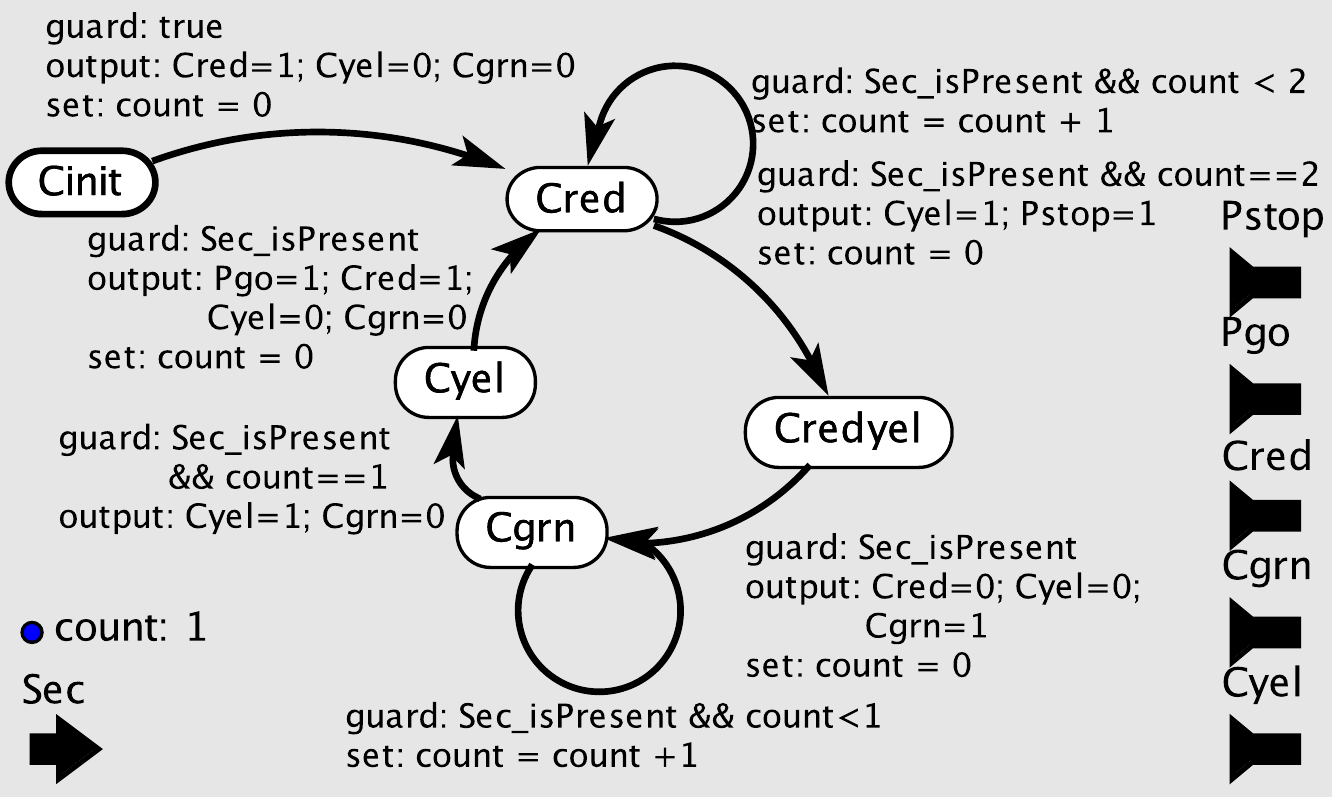}}                
  \caption{The FSM actors for pedestrian lights and car lights.}
  \label{fig:pedcar}
\end{figure}

Figure~\ref{fig:tl2} shows the FSM actor @PedestrianLight@. This actor has three
input ports (@Pstop@, @Pgo@, and @Sec@), two output ports (@Pgrn@ and @Pred@), three internal states,
and three transitions. This actor  reacts to signals from the 
car light (by way of the delay actors) by turning the pedestrian lights on and off. 
For example, if the actor is in local state @Pred@ and receives 
input through its @Pgo@  port, then it goes to state @Pgreen@, outputs the value 0 through its
@Pred@  port, and outputs the value 1 through its @Pgrn@ port.

Figure~\ref{fig:tl3} shows the FSM actor @CarLight@. 
Assuming that the \emph{clock} actor 
sends a signal every time unit, we notice, e.g., that one time unit
 after both the red and yellow car lights are on, 
these are turned off and the green car light is turned on by sending the
 appropriate values
 to the variables ({\small\texttt{output: Cred = 0; Cyel = 0; Cgrn = 1}}). 
The car light then stays green for 
 two time units before turning yellow. 

\subsection{Rewriting Logic and Real-Time Maude}
\label{sec:rtm}

A Real-Time Maude \emph{timed module} specifies a  \emph{real-time
rewrite theory}  of the form
 $(\Sigma, E, \mathit{IR}, \mathit{TR})$, where:
\begin{itemize}
\item  
$(\Sigma, E)$ is a \emph{membership equational
logic}~\cite{maude-book} 
theory with $\Sigma$ a signature\footnote{That is, $\Sigma$ is a set
  of declarations of \emph{sorts}, \emph{subsorts}, and
  \emph{function symbols}.} and $E$ a set of {\em confluent and terminating
conditional equations}. 
$(\Sigma, E)$
specifies
the system's state space as an algebraic data type, and must 
  contain a specification of a sort @Time@ modeling the (discrete or dense)
time
domain. 

\item  $\mathit{IR}$
is a set of (possibly conditional) \emph{labeled instantaneous rewrite
  rules} specifying  
the system's \emph{instantaneous} (i.e., zero-time) local transitions,  written 
$@rl [@l@] : @t@ => @t'$,
where  $l$ is a \emph{label}. 
%% Such a rule specifies a \emph{one-step transition}
%% from an instance of $t$ to the corresponding 
%% instance 
%% of $t'$. 
 The rules are applied \emph{modulo} the
equations~$E$.\footnote{$E = E'\cup
  A$, where $A$ is a set of equational axioms such as associativity,
  commutativity, and identity, so that deduction is performed \emph{modulo}
  $A$. Operationally,  a
term is reduced to its
$E'$-normal form modulo $A$ before any rewrite rule is applied.}

\item $\mathit{TR}$ is a set of \emph{tick (rewrite) rules}, written
  \quad  \texttt{rl [\(l\)]\!\! :\!\!\! \char123\(t\)\char125{}  => \char123\(t'\)\char125{} in time \(\tau\)},\;
that model time elapse.  @{_}@ is a 
built-in
 constructor of  sort \texttt{GlobalSystem}, and
$\tau$ is a term of sort @Time@ that denotes the \emph{duration}
of the rewrite.
\end{itemize}
The initial state must be a ground term of sort @GlobalSystem@ and 
 must be reducible to a term of
the form @{@$t$@}@ using the equations in the specifications. 

The Real-Time Maude syntax  is fairly intuitive. For example, 
 function symbols, or \emph{operators}, 
 are declared with the syntax \texttt{op }$f$ @:@ $s_1$ \ldots $s_n$
 @->@ $s$. $f$ is the name of the
 operator;
$s_1\:\ldots\:s_n$ are the sorts of the arguments of $f$; and $s$ 
 is its (value) \emph{sort}. Equations are written
with syntax @eq@ $t$ @=@ $t'$, and @ceq@ $t$ @=@ $t'$ @if@ \emph{cond}
for conditional equations. The mathematical variables in such statements
are declared with the keywords {\tt var} and {\tt vars}.
We refer to~\cite{maude-book} for more details  on the syntax of
 Real-Time Maude. 

We  use  the fact that 
an equation $f(t_1, \ldots, t_n) = t$ 
with  the @owise@ (for ``otherwise'') attribute  can be applied to a subterm 
$f(\ldots)$ only if no other equation with left-hand side
$f(u_1, \ldots, u_n)$ can be applied.\footnote{A specification with 
\texttt{owise} equations can be transformed to an equivalent system without
such equations~\cite{maude-book}.}

In object-oriented Real-Time Maude modules, a \emph{class} declaration

\small
\begin{alltt}
  class \(C\) | \(\attone\) : \(\sone\), \dots , \(\attn\) : \(\sn\) .
\end{alltt}
\normalsize

\noindent declares a class $C$ with attributes $att_1$ to $att_n$ of
sorts $s_1$ 
to $s_n$. An {\em object\/} of class $C$ in a  given state is
represented as a term
$@<@\: O : C \mid att_1: val_1, ... , att_n: val_n\:@>@$
of sort @Object@, where $O$, of sort @Oid@,  is the
object's
\emph{identifier}, and where $val_1$ to 
$val_n$ are the current values of the attributes $att_1$ to
$att_n$.
 In a concurrent object-oriented
system, the 
 state
 is a term of 
the sort @Configuration@. It  has 
the structure of a  \emph{multiset} made up of objects and messages.
Multiset union for configurations is denoted by a juxtaposition
operator (empty
syntax) that is declared associative and commutative, so that rewriting is 
\emph{multiset
rewriting} supported directly in Real-Time Maude.
The dynamic behavior of concurrent
object systems is axiomatized by specifying each of its 
transition patterns by a rewrite rule. For example, 
  the rule

{\small
\begin{alltt}
rl [l] : m(O,w)  < O : C | a1 : x, a2 : O', a3 : z >   =>
                 < O : C | a1 : x + w, a2 : O', a3 : z >  m'(O',x) .
\end{alltt}
}

\noindent  defines a  family of transitions 
%% (one for each substitution instance)  
in which a message @m@, with parameters @O@ and @w@, is read and
consumed by an object @O@ of class @C@. The transitions have the 
 effect of altering
the attribute @a1@ of the  object @O@ and of sending a new message
@m'(O',x)@.  
``Irrelevant'' attributes (such as @a3@)
need not be mentioned in a rule.

A \emph{subclass} inherits all the attributes and rules of its 
superclasses.

\paragraph{Formal Analysis.} 
A Real-Time Maude specification is \emph{executable}, and the tool offers a variety of formal analysis 
methods. The \emph{rewrite} command @(trew @$t$@ in time <= @$\tau$@ .)@
simulates \emph{one}
behavior of the system from initial state $t$ \emph{up to  duration} $\tau$.
The \emph{search}
command uses a breadth-first strategy to  analyze all possible
  behaviors of the system, by checking whether a state matching a
  \emph{pattern} and satisfying
  a \emph{condition} can be reached from the
  initial state.
%% The  command which searches for \emph{one} state
%% satisfying the search criteria has syntax
%% \begin{alltt}
%% (tsearch [1] \(t\) =>* \(pattern\) such that \(cond\) in time <= \(\tau\) .)
%% \end{alltt}

Real-Time Maude also extends Maude's \emph{linear temporal logic model
  checker} 
  to check whether
each behavior, possibly  up to a certain time bound,
  satisfies a temporal logic 
  formula.
 \emph{State propositions} are terms of sort @Prop@, and their 
semantics should be 
given by (possibly conditional) equations of the form
\texttt{\char123\(\mathit{statePattern}\)\char125{} |= \(\mathit{prop}\) = \(b\)}, 
for $b$ a term of sort @Bool@, which defines the state proposition 
$prop$ to hold in all
states $@{@t@}@$ where $@{@t@}@$ \verb+|=+ $prop$ evaluates to @true@.
A temporal logic \emph{formula} is constructed by state
propositions and
temporal logic operators such as @True@, @False@, @~@ (negation),
@/\@, @\/@, @->@ (implication), @[]@ (``always''), @<>@
(``eventually''), and @U@ (``until'').
The time-bounded model checking command has syntax 
\texttt{(mc \(t\) |=t \(\mathit{formula}\) in time <= \(\tau\) .)}
for initial state $t$ and temporal logic formula $\mathit{formula}$ .

    %% intro into Maude and Real-Time Maude
\section{Overview of the Formal Semantics of Flat Ptolemy DE Models} 
\label{sec:flat-ptolemy-semantics}

This section gives a brief overview of the Real-Time Maude
formalization of non-hierarchical and non-modal  (i.e., \emph{flat}) 
Ptolemy DE models given  in our paper~\cite{icfem09}. 
The reason for including  a summary of~\cite{icfem09} is:
(i) to convey the main idea of our semantics in a much simpler setting;
and (ii) to explain  why the semantics must be 
significantly changed for the hierarchical case.
To avoid
introducing too much  detail, we present a slightly simplified version
of our semantics, in that we, throughout the paper, 
 assume that all Ptolemy expressions are \emph{constants}.

%% \subsection{Supported Ptolemy Subset}
%% \label{sec:subset}
%%
%% \marginpar{MOVE elsewhere!}
%% 
%% We currently support Real-Time Maude analysis of   \emph{transparent}
%% \emph{discrete event} (DE) Ptolemy  
%% models;  that is, DE models where subdiagrams are also executed under
%% the DE director. We support composite actors, modal models,
%%  and the following  atomic actors:
%% finite state machine (FSM), timed delay, variable delay, clock, current
%% time, timer, noninterruptible timer, pulse, ramp, timed plotter, set
%% variable, and single event actors. We also support connections with
%% multiple destinations,
%% split 
%% signals, and both single ports and multi-input ports.

\subsection{Representing Flat Ptolemy DE Models in Real-Time Maude}
\label{sec:flat-semantics}

A \emph{flat} Ptolemy model is represented
as an object-oriented  Real-Time Maude term 

\small
\begin{alltt}
 \texttt{\char123}\(\mathit{actors}\)  \(\mathit{connections}\)  <\! global\! :\! EventQueue\! |\! queue\! : \(\mathit{event queue}\) >\texttt{\char125}
\end{alltt}
\normalsize

\noindent where
 \emph{actors} are objects corresponding to the actor instances in the
  Ptolemy model; 
\emph{connections} are the connections between the ports of the
  different actors; and where the value $\mathit{event\;queue}$ of the @queue@ attribute in the object
 \texttt{<\! global\!\! :\!\!\!\! EventQueue\!\! |\!\! queue\!\! :\!\!\!\! \(\mathit{event\;
      queue}\)\!\! > }
 denotes the global event queue.

\paragraph{Actors.}
Each Ptolemy actor is modeled  as an object instance
of a subclass of the following class @Actor@:

\small
\begin{alltt}
class Actor | ports : Configuration,  parameters : ValueMap .              
\end{alltt}
\normalsize

\noindent The @ports@ attribute denotes the set of \emph{ports} of the
actor. A port is modeled as an object, as shown below. The
@parameters@ attribute 
represents the \emph{parameters} of the corresponding Ptolemy actor,
together with their   values, as a
semicolon-separated set of 
terms of the form  @'@\textit{parameter-name} @|->@
$value$.

A \emph{timed delay} actor propagates an incoming event
after a given time delay. If the \emph{delay} parameter is
0.0, then there is a ``microstep'' delay on the generation of the
output event. Since the \emph{delay} parameter is represented
 in the @parameters@ attribute of @Actor@,  
this subclass does not add any attributes:

\small
\begin{alltt}
class Delay .        subclass Delay < Actor .
\end{alltt}
\normalsize

A \emph{finite state machine (FSM)  actor} is a transition system containing  finite sets of states (or ``locations''),
 local variables,  and
 transitions. A transition has 
a guard expression,  and can
contain a set of output actions  and variable assignments.  When an
FSM actor is 
fired, Ptolemy assumes that there is at most  one enabled transition. If there is an  enabled
transition then the actions in  
the transition are executed. Under the DE director, only one
transition step is performed in each iteration. 
An @FSM-Actor@ is characterized by its \emph{current state}, 
its  transitions, and the  values of its local variables:

\small
\begin{alltt}
class FSM-Actor | currState : Location,   initState : Location,
                  variables : ValueMap,   transitions : TransitionSet .
subclass FSM-Actor < Actor .
\end{alltt}
\normalsize

\noindent We model the transitions
as a semi-colon-separated set of
transitions of the form \newline{\small\texttt{\(s_1\!\) --> \(\!s_2\!\) \char123guard: \(\!\!g\!\) output:
  \(\!\!p_{i_1}\)\!|\!->\(\,e_{i'_1}\);\(\ldots\); \(\!p_{i_k}\)|\!->\(\,e_{i'_k}\!\)
  set: \(\!v_{j_1}\)|\!->\(\,e_{j'_1}\);\(\ldots\); 
\(\!v_{j_l}\)|\!->\(\,e_{j'_l}\)\char125}}
\newline for state/locations $s_1$ and $s_2$, port names $p_i$, variables $v_i$, and expressions
$e_i$.  

\paragraph{Ports and Connections.} 
A \emph{port} is modeled as an object, with a name (the identifier
of the object), a status (@unknown@, @present@, or @absent@), and
a @value@. We  have subclasses for input and output
ports:

\small
\begin{alltt}
class Port | status : PortStatus, value : Value .
class InPort .      class OutPort .      subclass InPort OutPort < Port .

sort PortStatus .   ops unknown present absent : -> PortStatus [ctor] .
\end{alltt}
\normalsize

 A  \emph{connection} is represented as  a term 
$p_{o}$ @==>@ $p_{i_1};\ldots ; p_{i_n}$ of sort
@Connection@, where each $p_j$ has the
form $a @!@ p$ for $a$ the name of an actor and $p$ the  name of a port. Such a connection 
connects the output port $p_o$ to all the input ports $p_{i_1}, \ldots, 
p_{i_n}$
Since connections appear in configurations, the sort @Connection@ is defined to be 
a subsort of
the sort @Object@.

\paragraph{Global Event Queue.} The global event queue is
maintained by an object @global@ of class @EventQueue@ whose @queue@ attribute 
represents the global event queue as a @::@-separated list, ordered according to time  until firing,
 of terms of
the form\quad \texttt{\(\mathit{set\,of\,events}\) ; \(\mathit{time\,to\,fire}\) ; \(\mathit{microstep}\)}. 
%% \begin{alltt}
%%  \(\mathit{set\,of\,events}\) ; \(\mathit{time\,to\,fire}\) ; \(\mathit{microstep}\)
%% \end{alltt}
The \emph{set of events} is a set of events, each event
characterized by the ``global port name'' where the generated event
should be output and the corresponding value; \emph{time to fire}
denotes the time \emph{until} the events are supposed to fire; and
\emph{microstep} is the additional ``microstep'' until the event
fires.

\paragraph{Example: Representing the Flat Traffic Light Model.}

 The Real-Time Maude representation
of the @TimedDelay2@ delay actor in the flat non-fault-tolerant traffic light system  in
Section~\ref{sec:pt-example} is 

\footnotesize
\begin{alltt}
< 'TimedDelay2 : Delay | parameters : 'delay |-> # 1.0,
                         ports : < 'input : InPort | value : # 0, status : absent >
                                 < 'output : OutPort | value : # 0, status : absent >  >
\end{alltt}
\normalsize

\noindent Likewise, the FSM actor
@CarLightNormal@ in the initial state 
is represented as the term\footnote{To save space, some terms are replaced by `\texttt{...}'} 

\footnotesize
\begin{alltt}
< 'CarLight : FSM-Actor | initState\! : 'Cinit,  currState\! : 'Cinit,  variables\! : 'count |-> # 1,
                          ports : <\! 'Sec : InPort | value : # 0, status : absent\! >
                                  <\! 'Pgo : OutPort | value : # 0, status : absent\! >  ..., 
                          transitions : ('Cinit --> 'Cred 
                                           \char123guard: (# true) 
                                            output: ('Cred\! |->\! #\!\! 1)\!\! ;\!\! ('Cyel\! |->\! #\!\! 0)\!\! ;\!\! ('Cgrn\! |->\! #\!\! 0)
                                            set: 'count |-> # 0\char125) ;
                                        ('Cred --> 'Cred 
                                           \char123guard: (isPresent('Sec) && ('count lessThan # 2)) 
                                            output: emptyMap  
                                            set: 'count |-> ('count + #\!\! 1)\char125) ; ...\! >\! .
\end{alltt}
\normalsize

The connection from the output port @output@ of the @Clock@ actor to the input port @Sec@ of
@CarLight@  and the input port @Sec@ of @PedestrianLight@ is
represented by the  term

\footnotesize
\begin{alltt}
(\!'Clock\!\! !\!\!\! 'output)\! ==>\!\! (\!'PedestrianLight\!\! !\!\! 'Sec)\!\! ;\!\! (\!'CarLight\!\! !\!\! 'Sec)
\end{alltt}
\normalsize

%% The entire state thus consists of two FSM actor objects, ten\marginpar{remove?}
%% connections, two delay objects, five SetVariable objects, and the
%% global event queue object.

\subsection{Specifying the Behavior of Flat DE Models}

As explained in Section~\ref{sec:ptolemy-intro}, 
the behavior of Ptolemy DE models can be summarized as repeatedly:
\begin{itemize}
\item Advance time  until the time to fire 
  the first events in the  queue is $(0,0)$.
\item And then perform an iteration of the system. That is:
\begin{enumerate}
\item The events that are supposed to fire are added to
  the corresponding output ports; the @status@ of all other ports is
  set to @unknown@.
\item (Fire) Then the \emph{fixed point} of all ports is computed by
  gradually increasing the knowledge about the presence/absence of
  inputs to and output from ports until a fixed-point is reached.
\item (Postfire) Finally, states are updated for  actors with inputs or scheduled events, 
 and new events are generated and
  inserted into the event queue.
\end{enumerate}
\end{itemize}

The following  tick rule advances time until the time when the
first events in the event queue are scheduled (we first declare all the variables used):

\small
\begin{alltt}
vars SYSTEM OBJECTS PORTS PORTS' REST REST2 IA ACTS\! : ObjectConfiguration\!\! . var N\! :\! Nat\!\! . 
var EVTS\! : Events\! . var QUEUE\! : EventQueue\! . vars V TV\! : Value . vars PARAMS\! :\! ValueMap .
vars T T'\! : Time\! . var NZT\! : NzTime\! . vars STATE STATE'\! : Location . vars O O' CO : Oid\!\! .
var CF\! :\! Configuration\! . var BODY\! : TransBody\!\! . var TRANSSET\! : TransitionSet\!\! . 
vars P P'\! : PortId\! . var EPIS\! : EportIdSet\! . var PS\! : PortStatus\! . var NZ\! : NzNat\! . 
\end{alltt}
\begin{alltt}
rl [tick] : 
   \texttt{\char123}SYSTEM  <\! global\! :\! EventQueue\! |\! queue\! : (EVTS\! ;\! \emph{NZT}\! ;\! N) :: QUEUE\! >\texttt{\char125}
  => 
   \texttt{\char123}delta(SYSTEM, NZT) 
    <\! global\! :\! EventQueue\! |\! queue : (EVTS\! ;\! \emph{0}\! ;\! N) :: delta(QUEUE, NZT)\! >\texttt{\char125}  \emph{in time NZT}\! .
\end{alltt}
\normalsize

\noindent The first event(s) in the event queue have
non-zero delay 
@NZT@. Time is advanced by this amount @NZT@, and, as a consequence,
the (first component of the) event timer goes to zero. In addition, the
function @delta@ is applied to all the other objects (denoted by
@SYSTEM@) in the
system. The function @delta@ defines the effect of time elapse on the
objects. This function is also applied to the other elements in the
event queue, where it decreases the remaining time of each event set
by the elapsed time @NZT@ (see~\cite{icfem09} for details).

The   ``microstep tick rule'' that advances ``time'' 
by microsteps is not shown. 
When the remaining time and microsteps of the first events
in the  queue are both zero, an
iteration of the system is performed:

\small
\begin{alltt}
rl [executeStep] : 
   \texttt{\char123}SYSTEM < global : EventQueue | queue : (EVTS ; 0 ; 0) :: QUEUE >\texttt{\char125}
  =>    
   \texttt{\char123}< global : EventQueue | queue : QUEUE >
    postfire(portFixPoints(addEventsToPorts(EVTS, clearPorts(SYSTEM))))\texttt{\char125}\!\! .
\end{alltt}
\normalsize

\noindent The function @clearPorts@ 
sets the @status@ of each port 
 to @unknown@. The operator @addEventsToPorts@
inserts the events scheduled to fire into the corresponding output
ports.  The  
@portFixPoints@ function then finds the fixed points for all  ports
(fire), and @postfire@ ``executes'' the steps on the computed port
fixed-points by changing the states of the
objects and generating new events and inserting them into the global event
queue. These functions have sort @Configuration@, whereas  the equations defining them
 involve variables of the subsort \texttt{Object\-Configuration},  so
that  each function has finished computing before the ``next'' function is computed:

\small
\begin{alltt}
ops clearPorts portFixPoints postfire : Configuration ~> Configuration .
\end{alltt}
\normalsize 

To completely define the behavior of a system, we must define
 @clearPorts@,  
@portFixPoints@, and @postfire@ on the different actors.

%% \paragraph{Initialize Actors.}
%% In our simplified setting, the @clearPorts@ function just clears
%% all the ports, that is, sets their @status@ to @unknown@:
%% 
%% \small
%% \begin{alltt}
%% eq clearPorts(< O : Actor | ports : PORTS > REST) 
%%      = < O : Actor | ports : clearPorts(PORTS) >   clearPorts(REST) .
%% eq clearPorts(SYSTEM) = SYSTEM [owise] .
%% 
%% op clearPorts : Configuration -> Configuration .
%% eq clearPorts(< P : Port | status : PS > PORTS)
%%      = < P : Port | status : unknown > clearPorts(PORTS) .
%% eq clearPorts(none) = none .
%% \end{alltt}
%% \normalsize

\paragraph{Computing the Fixed-Point for Ports.}

The idea behind the  function @portFixPoints@, that
computes
the fixed-point for  all  ports, is simple. The state has
the form @portFixPoints@(\emph{actors and connections}), where
initially, the only 
port information are the events scheduled for this iteration. For each
 case
when the status of an @unknown@  port can be determined to be either
@present@ or @absent@, there is an  equation

\small
\begin{alltt}
eq portFixPoints(<\! O\! :\! ...\! |\! ports\! : <\! P\! :\! Port\! |\! status\! :\! \emph{unknown} > PORTS,\, ... >
                 \(\mathit{connections and other objects}\)) 
 = portFixPoints(<\! O\! :\! ...\! |\! ports\! : <\! P\! :\! Port\! |\! status\!\! :\! \emph{present}\!\!,\! value\!\! :\! ...\! > PORTS,\! ...\! >
                 \(\mathit{connections and other objects}\)) .
\end{alltt}
\normalsize

\noindent (and similarly for deciding that input/output is
@absent@). The fixed-point is reached when no such
%% ``information-adding'' 
equation can be applied. The 
\texttt{portFixPoints}  operator is then removed by using the @owise@ construct:

\small
\begin{alltt}
eq portFixPoints(OBJECTS) = OBJECTS \emph{[owise]} .
\end{alltt}
\normalsize

The following equation propagates port status from a ``known'' output
port to a 
connecting @unknown@ input port. The present/absent @status@ (and
possibly the @value@) of the
output port @P@ of actor @O@ is propagated to the input port @P'@ of
the actor @O'@ through the connection 
\texttt{(O ! P) ==> ((O' ! P') ; EPIS)}:

\small
\begin{alltt}
ceq portFixPoints(
      < O : Actor | ports :  < P : OutPort | status : \emph{PS}, value : V > PORTS >
      ((O ! P) ==> ((O' ! P') ; EPIS)) 
      < O' : Actor | ports : < P' : InPort | status : \emph{unknown} > PORTS' > 
      REST)
  = portFixPoints(< O : Actor | >  ((O ! P) ==> ((O' ! P') ; EPIS)) 
      < O' : Actor | ports : < P' : InPort | status : PS, value : V > PORTS' >
      REST)   \emph{if PS =/= unknown} .
\end{alltt}
\normalsize

The @portFixPoints@ function must then be defined for each kind of
actor to decide whether the actor produces any output in a given
port. 
For example, the \emph{timed delay} actor does not produce any output
in this iteration as a result of any input. Therefore, if its @status@
is @unknown@ (that is, the delay actor did not schedule an event for
this iteration), its output port should be set to @absent@:

\small
\begin{alltt}
eq portFixPoints(< O : Delay\! |\! ports\! : < P\! :\! OutPort\! |\! status\! :\! \emph{unknown} > PORTS > REST)
 = portFixPoints(< O : Delay\! |\! ports\! : < P\! :\! OutPort\! |\! status\! :\! \emph{absent} > PORTS > REST)\! .
\end{alltt}
\normalsize

The definition of @portFixPoints@ for FSM actors relies on the assumption that at most one 
transition is enabled at any time. In the following conditional equation, 
one  transition from the current state @STATE@ is enabled. In addition, there is \emph{some}
input to the actor (through input port @P'@), and some output ports have status @unknown@.
 The function @updateOutPorts@ then updates the status and the  values of the output
ports according to the current state and input:

\small
\begin{alltt}
ceq portFixPoints(< O\! :\! FSM-Actor | ports : < P' : InPort | status : present >
                                           < P : OutPort | status : unknown > PORTS,
                                   currState : STATE, parameters : PARAMS,
                                   transitions\! :\! (STATE --> STATE'\! \char123BODY\char125)\!\! ;\!\! TRANSSET\! >
                   REST)
  = portFixPoints(< O\! :\! FSM-Actor | ports : updateOutPorts(PARAMS, BODY,
                                              < P\!\! :\!\! OutPort\! |\! > < P'\!\! :\!\! InPort\! |\! > PORTS)\! >
                   REST) 
 if transApplicable(< P\! :\! OutPort\! |\! > < P'\! :\! InPort\! |\! > PORTS, PARAMS, BODY) .
\end{alltt}
\normalsize

\noindent Another equation sets all output ports to @absent@ if there is
enough information to determine that no transition can become
 enabled in the 
current round.

\paragraph{Postfire.} The @postfire@ 
function  distributes over the 
actor objects in the configuration and updates
 internal states and generates new events that are inserted
 into the event queue.  An \textit{owise} equation 
defines  @postfire@  to be the identity function 
 on those
actors that do not have other equations defining @postfire@.

If a  \emph{delay}
 actor has input in its
@'input@ port, then it generates an event with a delay
equal to the value of the @'delay@ parameter. 
If this delay is @0.0@, the microstep is @1@, otherwise the microstep
is @0@. This event is added to the global event queue using the @addEvent@ function 
that adds the new event to the  queue:

\small
\begin{alltt}
eq postfire(< O\! :\! Delay\! |\! ports : < 'input\! :\! InPort\! |\! status : present, value : \emph{V} >  
                                 \!< 'output\! :\! OutPort\! |\! >, 
                         parameters : 'delay |-> \emph{TV} ; PARAMS >)
   < global\! :\! EventQueue\! |\! queue : QUEUE >
  = 
   <\! O\! :\! Delay\! |\! >
   <\! global\! :\! EventQueue\!\! |\!\! queue\! :\! addEvent(event(O\! !\! 'output,\! V)\!,\! toTime(TV)\!,
                             \!   \!  if toTime(TV)\! ==\! 0\! then\! 1\! else\! 0\! fi,\! QUEUE)\!\! >\!\! .
\end{alltt}
\normalsize

An FSM actor does not generate future events, but
@postfire@ updates the location and variables of the
actor if it has input and has an enabled transition; again, the rule is given in~\cite{icfem09}.

\section{Hierarchical Ptolemy DE Models} 
\label{sec:hierarchical-ptolemy}

Ptolemy II  \emph{hierarchical} models contain components (or
\emph{actors})
that are themselves Ptolemy II models. 
Such a hierarchical model can again be encapsulated and be seen as a single
\emph{composite actor}
(Non-composite actors are called {\em atomic actors}).
%% This helps hide internal details of parts of a model, and therefore is
%% crucial for managing model complexity, and for achieving modularity and
%% scalability. 
An inner actor of a DE composite actor is
executed if that inner actor receives some events at its input ports.  
Ptolemy II also provides \emph{modal models}
where several sub-models are controlled by a state machine.
%where each state of the machine can be refined into an internal model.
Each ``state'' of a modal model is (or ``refines to'') a Ptolemy
model, and only the model of the current state is
executed during the computation step.

\begin{figure}[htb]
\begin{center}
\includegraphics[scale=0.72]{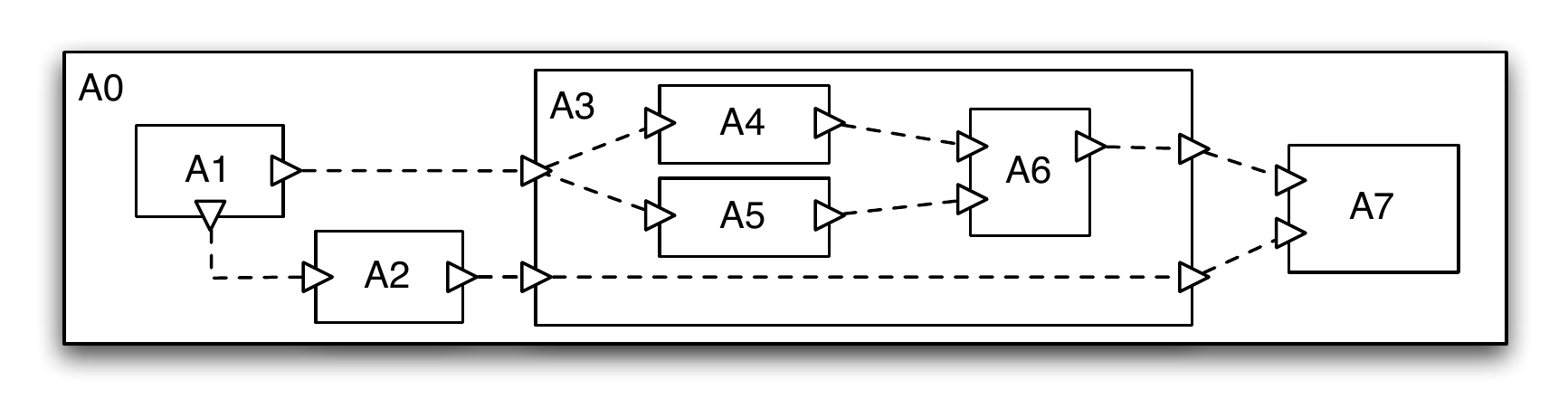}
\caption{A hierarchical composition of actors.
$A0$--$A7$ are actors, and $A0$ and $A3$ are composite actors.}
 \label{fig:hie-diag}
 \end{center}
\end{figure}

\paragraph{Composite Actors.}
Composite actors can have parameters, ports to
communicate with  other actors, and  a sub-model that can be any Ptolemy model. 
The ports of a composite actor are connected to its inner actors so
that the sub-model interacts with 
the outside.
%% At the same level, the topology of connections
%% are identical with the flat case, but 
%% Ports of composite actors interact differently with 
%% inner actors.
The input ports of composite actors are connected
to the input ports of inner actors, and the output ports of composite actors are connected to
the output ports of inner actors.
Figure~\ref{fig:hie-diag} illustrates a hierarchical composition of actors.
%% with connections in composite actors.

In Ptolemy II, each composite actor can have its own \emph{director}
to support heterogeneous modeling. If the director of a composite actor
is the same as the director of the parent actor, it is called a \emph{transparent} actor. 
In this paper, we consider only transparent cases since we verify  DE models.
The evaluation order of actors in a transparent hierarchical model is essentially 
the same as in  the flat case. 
A composite actor is \emph{fired} if a new value has arrived to
an input port of the actor, and the value is transferred to the ports connected to the
input port. If an output port of a composite actor gets a new value, either from inner actors
or from input ports of the  actor, the value is transferred to the connected ports.
This \emph{port fixed-point} computation  finishes when no other port-transfer is 
available, just as in  the flat case. 

\begin{figure}[htb]
\begin{center}
\includegraphics[scale=0.72]{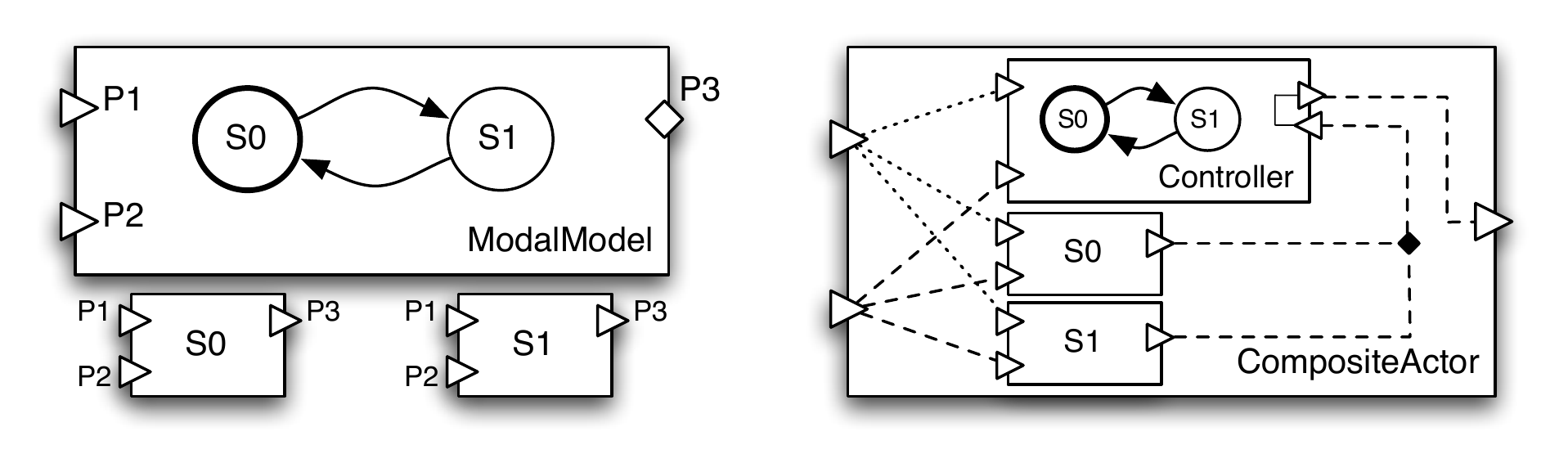}
\caption{A modal model with 2 states and its equivalent representation as a composite actor. 
$S0$ and $S1$ are states, triangles are ports, and diamonds are input/output ports.
Dashed lines are connections, and a solid line in the right-hand side 
means a coupled input/output ports.}
\label{fig:modal-model}
\end{center}
\end{figure}

\paragraph{Modal Models.}
Modal models are  finite state machines where 
each state has a \emph{refinement} actor, which is either
a composite actor or an FSM actor. The input and output ports of the refinements are the same as
those of the modal model. 
In the top level of a modal model, the output ports are regarded
as \emph{both} input and output ports 
so that the transitions of modal models may use the evaluation result of 
refinement actors in the \emph{current} computation step.
%%% I removed the sentence below, hope it is OK.
%% Note that with only parameters modal models cannot use the evaluation result of 
%% refinements in the current step since each parameter is updated at the
%% end of each computation step (\texttt{postfire}). 
The left-hand side  of Fig.~\ref{fig:modal-model}  shows  a modal model 
with two states.

When a modal model fires, the refinement of the current state is fired and the other refinements
are \emph{frozen}. The guards of all
outgoing transitions from the current state of the modal model are then evaluated.\footnote{
Both the input ports and the input/output ports of modal models can be used here.}
If exactly one of those guards is true, then
the transition is taken and the actions on the transition are executed. 
If more than one of the guards is true, then it is considered as an error in Ptolemy II.
The refinement of the next state will be executed in the \emph{next} computation step.
In  case of a conflict between the refinements and the parent actor, the latter overwhelms the former. 
For example, if the FSM controller of a modal model and the refinement of a current state
are trying to write different values to the same output port, then the value of the FSM controller is taken.
%Indeed, Ptolemy II implementation works this way.

A modal model can be seen as  syntactic sugar for a  composite actor
with \emph{frozen} inner actors, as shown in Fig.~\ref{fig:modal-model}, 
where the right-hand side shows  the equivalent composite-actor
representation of the modal model in the left-hand side. 
That is, a modal model $A$ is semantically equivalent to  a composite actor $\tilde{A}$,
with the same ports, that has the controller FSM actor and the refinement actors as inner actors, so that:
(i) the ports are connected as indicated in Fig.~\ref{fig:modal-model}; 
(ii)  the controller FSM actor is fired \emph{after} the refinement actors are fired;
(iii) only the refinement inner actors corresponding to the current state of the controller are evaluated, whereas
                the other refinement  actors  are frozen,  
		in the sense that their states do not evolve and the values of their outports are ignored; and
	(iv) if an output port of the controller FSM actor has no value but its coupled input port has some value,
		then the output port will have the same value as the input port.
%%		Otherwise, the value of coupled input port can be used by a transition of the controller FSM actor,
%%		but it is not transferred to $\tilde{A}$.
%%% Kyungmin, all these bullets take too much space. I am trying to rewrite this.
%%% Also, how important are all these details for our purpose??
%% \begin{itemize}
%% 	\item $\tilde{A}$ has the same set of ports as $A$.
%% 	\item $\tilde{A}$ has the controller FSM actor and the refinement actors as inner actors,
%%		where all those inner actors have the same sets of input and output ports as $A$.
%%	\item Each output port of the controller FSM actor has a \emph{coupled} input port,
%%		since  the output port of a modal model are considered as both an input and output port.
%%	\item Each port of inner actors is connected to its original copy in $\tilde{A}$,
%%		from an input port to an input port, and from an output port to an output port.
%%	\item The controller FSM actor is fired after the refinement actors are fired.
%%	\item Only the refinement inner actors corresponding to the current state of the controller are evaluated.
%%                The other refinement inner actors  are frozen, 
%%		in the sense that their states do not evolve and the values of their outports are ignored.		
%%	\item If an output port of the controller FSM actor has no value but its coupled input port has some value,
%%		then the output port will have the same value of the input port.
%%		Otherwise, the value of coupled input port can be used by a transition of the controller FSM actor,
%%		but it is not transferred to $\tilde{A}$.
%% \end{itemize}
Our Real-Time Maude semantics in Section \ref{sec:hierarchical-semantics}
follows this semantics.

\section{Real-Time Maude Semantics for Hierarchical DE Models}
\label{sec:hierarchical-semantics}

We define the Real-Time Maude semantics for hierarchical DE models 
by extending our semantics for flat models to 
composite actors and modal models, and by making
some changes to the flat semantics as described below. 
Our Real-Time Maude representation  preserves the hierarchical structure of a Ptolemy II model;
 therefore such  models and their Real-Time Maude counterparts are essentially isomorphic, so 
that we can easily reconstruct the original Ptolemy II models to provide
 graphical counter-examples.

Some of the difficulties involved in  extending the semantics   to the hierarchical case 
 include: 
\begin{itemize}
\item The event management is different. DE models have a \emph{global}  event queue, but events 
   could be generated at  any level and may need to be delivered deep down in the hierarchy.
\item  Computing fixed-points for hierarchical models is much harder than in the flat case. 
  Naive approaches easily fall into infinite loops or unnecessarily complex semantics.
  In addition, the fixed-point computation should be finished only after 
all levels of  fixed-point computation are completed.
\item  The  semantics of modal models in the Ptolemy II documentation is somewhat  unclear.
  There are many subtle or implicit assumptions concerning the execution of modal models,
  such as the evaluation order of inner actors, event generation in frozen actors, and
  handling input/output ports of modal models.
  We proposed the transformation from modal models  to composite actors
  for clarifying the semantics of modal models, and have discussed this issue extensively 
with members of the Ptolemy team. 
\end{itemize}

\subsection{Representing Hierarchical Actors}
\emph{Composite actors} are modeled as
object instances of the  class @CompositeActor@, which extends its superclass @Actor@ with 
  one attribute, @innerActors@, which denotes the 
inner actor objects and connections of the composite actor:

\small
\begin{alltt}
class CompositeActor | innerActors : Configuration .
subclass CompositeActor < Actor .
\end{alltt}
\normalsize

We also add the following new class @AtomicActor@, and 
 declare each \emph{atomic} actor
class to be a subclass of @AtomicActor@. 

\small
\begin{alltt}
class AtomicActor .    subclass AtomicActor < Actor .
\end{alltt}
\normalsize

Each actor can be uniquely identified by its
 \emph{global actor identifier}, which  is a list \texttt{\(o_1\,\).\(\,o_2\,\).\(\;\ldots\;\).\(\,o_n\)} of object names,
where $o_1$ is the name a top-level actor, and which 
includes all identifiers of composite actors containing the given actor.

We represent modal models as  composite actors according
to the frozen-composite-actor semantics for modal models described in 
Section~\ref{sec:hierarchical-ptolemy}. 
The class @ModalModel@ has an additional attribute 
@controller@ pointing to the controller FSM in @innerActors@, and 
the additional @refinementSet@ attribute mapping each state 
in the modal model to its refinement:

\small
\begin{alltt}
class ModalModel | controller : Oid,  refinement : RefinementSet .
subclass ModalModel < CompositeActor .
\end{alltt}
\normalsize

In addition,
the definition of the basic @Actor@ class adds an attribute
@status@ which can have the value @enabled@ or @disabled@ to reflect
that  any actor may be disabled as a result of being contained in a refinement
of a state in a modal model which is ``frozen.'' 
The equations for @postfire@ and @portFixPoints@ generating a value at outports
only apply to objects whose @status@ is @enabled@. 
The other equations such as @clearPorts@ are applied to @disabled@ actors.

\subsection{Extracting and Adding Event to the Event Queue}

In the flat setting, each actor is at the same hierarchical level
as the  global @EventQueue@ object. Each actor therefore has direct access
to the event queue, so that at the start  of an iteration, the scheduled events
could be directly inserted into the corresponding actor ports
(by the function  @addEventsToPorts@), and actors could add generated events
directly into the global event queue (by @postfire@). 

In the hierarchical case, an actor that receives or generates an event
from/to the global event queue can be located deep down in the 
actor hierarchy. Events communicated between the actors and the  event
queue may therefore cross hierarchical boundaries. 
We have modeled this
``traveling'' of events by ``method calls'' or ``messages''. For example, 
inserting an event into the output port $p$ of some actor with
global actor identifier $g$ corresponds to generating the message
\texttt{active-evt(event(\(g\:\)!\(\:p\),\(\:v\)))}. 
Likewise, 
an event generated by an  actor is ``sent'' to the event queue as a ``message''
of  the form
\texttt{schedule-evt(\(\mathit{event}\),\(\,\mathit{time}\),\(\,\mathit{microstep}\))}:

\small
\begin{alltt}
msg schedule-evt : Event Time Nat -> Msg .
msg active-evt : Event -> Msg .
\end{alltt}
\normalsize

For example, when an actor generates an event, it creates a
@schedule-evt@ ``message'':

\small
\begin{alltt}
eq postfire(< O : Delay | status : enabled,  parameters : 'delay |-> TV ; PARAMS,
                          ports : < 'input : InPort | status : present, value : V >  
                                  < 'output : OutPort | > PORTS >)
 = \emph{schedule-evt(event(O\! !\! 'output,\! V),\! toTime(TV), if toTime(TV) == 0 then 1 else 0 fi)} 
   < O : Delay | > .
\end{alltt}
\normalsize

\noindent Such an event is propagated  towards the top of the actor
hierarchy by the following equation, in which a composite actor
inside whose @innerActors@  the @schedule-evt@ message
resides moves the message one level up:

\small
\begin{alltt}
eq < O : CompositeActor | innerActors : CF \emph{schedule-evt(event(AI ! PI, V), T, N)} >
 = < O : CompositeActor | innerActors : CF >
   \emph{schedule-evt(event((O . AI) ! PI, V), T, N)} .
\end{alltt}
\normalsize

\noindent When the @schedule-evt@ request has reached
the top  of the hierarchy, it is added to  the event queue:

\small
\begin{alltt}
eq < global : EventQueue | queue : QUEUE > \emph{schedule-evt(EVENT, T, N)}
 = < global : EventQueue | queue : addEvent(EVENT, T, N, QUEUE) > .
\end{alltt}
\normalsize

 The propagation of @active-evt@s from the event queue to some inner actor is explained below.

The rewrite rule @executeStep@ that models  an iteration of the system is modified w.r.t.\ the flat case,
 so that for each event 
\texttt{event(\(\mathit{globalActorId}\;\)!\(\;\mathit{portId}\),\(\;v\))} scheduled for this iteration (i.e., included 
 in \texttt{EVTS} below), a ``message'' 
\texttt{active-evt(event(\(\mathit{globalActorId}\;\)!\(\;\mathit{portId}\),\(\;v\)))} is added
to the state; the function @releaseEvt@ generates this message set from a set of events:

\small
\begin{alltt}
rl [executeStep] : 
   \char123SYSTEM  <\! global\! :\! EventQueue\! |\! queue\! : (EVTS ; 0 ; 0) :: QUEUE\! >\char125
  =>
   \char123<\! global\! :\! EventQueue\! |\! queue\! : QUEUE >
    postfire(portFixPoints(releaseEvt(EVTS) clearPorts(SYSTEM)))\char125 .
\end{alltt}
\normalsize

%% \noindent The functions @releaseEvt@, @prefire@, and @portFixPoints@  have sort @Configuration@.
%% Subtle use of variables of the subsort   \texttt{ObjectConfiguration} ensure that
%% @prefire@  has been computed  \emph{before}  @portFixPoints@ is computed, and that 
%% @portFixPoints@ is computed before @postfire@ is computed. 

\subsection{Defining \texttt{clearPorts}, \texttt{portFixPoints}, and \texttt{postfire} 
for Hierarchical Models}

For \emph{atomic} actors, @clearPorts@ should just set the @status@ of each port of the 
actor to @unknown@, as before. For \emph{composite} actors, it should also propagate to the inner actors.
 To ensure that the appropriate
equation applies to an actor, we must modify the definition of
@clearPorts@ for atomic actors to apply only to objects of class @AtomicActor@:

\small
\begin{alltt}
eq clearPorts(<\! O\! :\! AtomicActor\! |\! ports\! :\! PORTS\! >) 
 = <\! O\! :\! AtomicActor\! |\! ports\! :\! clearPorts(PORTS)\! > .
eq clearPorts(<\! O\! :\! CompositeActor\! |\! innerActors\! :\! CF, ports\! :\! PORTS\! >)
 = <\! O\! :\! CompositeActor\! |\! innerActors\! :\! \emph{clearPorts}(CF), ports\! :\! clearPorts(PORTS)\! > .
\end{alltt}
\normalsize

The @postfire@ function is almost unchanged for the ``flat'' actors; the only
modification is to ensure that @postFire@  is not  applied to  \emph{disabled} actors, since 
 disabled actors should not change their states or  generate new events. 
 For a composite actor, @postfire@ just
propagates to its  inner actors. The condition ensures that this equation
is not applied to modal models:

\small
\begin{alltt}
ceq postfire(<\! O\! :\! CompositeActor\! |\! status\! :\! ST, innerActors\! :\! CF\! >)
 = <\! O\!\! :\!\! CompositeActor\!\! |\!\! innerActors\! :\! if ST == enabled then \emph{postfire}(CF)\! else CF fi\! >)\! 
  if class(<\! O\!\! :\!\! CompositeActor\!\! |\!\! >) == CompositeActor .
\end{alltt}
\normalsize

The extension of the @portFixPoints@ function to the hierarchical case is more
subtle due to repeated computations.
The @portFixPoints@ function should distribute to the submodels of
composite actors, to compute the fixed points of these subsystems. 
However, an equation of the form 

\small
\begin{alltt}
eq portFixPoints(< O : CompositeActor | innerActors : IA, ... > REST) 
 = portFixPoints(< O : CompositeActor | innerActors : portFixPoints(IA), ... > REST) .
\end{alltt}
\normalsize

\noindent would be applicable again when the inner
@portFixPoints@ function disappears, leading to nontermination
(and non-applicability of the @owise@ equation defining the end of the
fixed-point computation). 
This anomaly is actually caused by applying
@portFixPoints@ to inner actors even though
they may already have reached their fixed points.
To avoid this situation, we need to execute @portFixPoints@ for inner actors
\emph{only if} some inner actors have not yet reached a fixed-point.
This can be easily accomplished since actors are activated in DE models only 
if input ports of the actors receive some value either from the event queue or from
the other actors by connections.

We therefore start the fixed-point computation of inner actors
in the @portFixPoints@ function of composite actors for the following cases:
 \begin{enumerate}
	\item Some events from the global event queue are passed to some inner actors.
	\item An input port of a composite actor connected
		to some inner actors receives some value.
\end{enumerate}

For Case 1,
when  released events in a configuration should be propagated to some 
inner actor of a composite actor, we  begin the @portFixPoints@ computation of those inner actors.
The following equations describe the propagation of @active-evt@s from the event queue
to inner actors. If there are some events toward an inner actor of a composite actor,
all such events are then passed to the inner actors and @portFixPoints@ of the
inner actors is started. This equation is the only equation defined on the sort @Configuration@
so that it is executed first before the other fixed-point equations are evaluated:

\small
\begin{alltt}
ceq portFixPoints(active-evt(event((O . AI) ! PI, V)) 
                  < O : CompositeActor | innerActors : ACTS > CF)
  = portFixPoints(< O : CompositeActor | innerActors : \emph{portFixPoints}(MSGS ACTS) > CF')
 if fr(MSGS, CF') := filterMsg(O, CF, active-evt(event(AI ! PI, V)) ) .
\end{alltt}
\normalsize

\noindent The function @filterMsg@ separates the events toward inside
from the others, and returns a constructor \texttt{fr(\(\mathit{Events}\),\(\mathit{Conf}\))}
which is a pair of the desired events and the other configuration.

%% \small
%% \begin{alltt}   
%% sort FilterResult .
%% op fr : MsgConfiguration Configuration -> FilterResult [ctor] .
%% op filterMsg : Oid Configuration MsgConfiguration -> FilterResult .
%% 
%% eq filterMsg(O, active-evt(event((O . AI) ! PI, V)) CF, MSGS)
%%  = filterMsg(O, CF, active-evt(event(AI ! PI, V)) MSGS) .
%% eq filterMsg(O, CF, MSGS) = fr(MSGS, CF) [owise] .
%% \end{alltt}
%% \normalsize

In Case 2, we must  define the @portFixPoints@ function for the port-propagation
of composite actors.
An input to a composite
actor will lead to an input to one of its subactors, and an output
at a subactor will lead to an output from the containing
composite actor (the special name `@parent@' denotes the 
containing actor of an actor in port names).
When a composite actor passes a value (or the knowledge that input will be @absent@) to inner actors,
if the inner fixed-point computation has not started yet or is already finished,
then @portFixPoints@ must again be called to (re-) compute the fixed-point of the 
inner diagram:

\small
\begin{alltt}
ceq portFixPoints(
        < O : CompositeActor |
                status : \emph{enabled},
                ports : < P : InPort | status : PS, value : V >  PORTS,
                innerActors :
                  (\emph{parent} ! P) ==> (O' ! P' ; EPIS)
                  < O' : Actor | ports : < P' : InPort | status : \emph{unknown} > PORTS2 >
                  REST2 >
        REST) 
  =   
    portFixPoints(
        < O : CompositeActor | 
                 innerActors : \emph{portFixPoints(}    *** (re-)\! start the inner fixed-point
                   (\emph{parent} ! P) ==> (O' ! P' ; EPIS)
                   <\! O'\! :\! Actor\! |\! ports\! :\! <\! P'\! :\! InPort\! |\! status\! :\! \emph{PS},\! value\! :\! V\! > PORTS2 >
                   REST2) >
        REST)  \emph{if PS =/= unknown} .
\end{alltt}
\normalsize

%% From journal paper
Of course, a composite actor can pass an updated port status/value to its inner actors also
when those inner actors are already computing @portFixPoints@; that case is modeled by an equation
that is very similar to the above equation and is not shown.

Likewise, an inner actor can propagate the status of output ports to the containing actor.
In this case, we only consider when the inner fixed-point is already finished,
because in Ptolemy II  an inner actor has a higher priority than a parent actor
in the evaluation order:

\small
\begin{alltt}
ceq portFixPoints(
       < O : CompositeActor | 
                ports : < P : OutPort | status : \emph{unknown} >  PORTS,
                innerActors : 
                  (O' ! P') ==> (\emph{parent} ! P ; EPIS)
                  < O' : Actor | status\! : \emph{enabled},
                                 ports\! : <\! P'\!\! :\! OutPort\! |\! status\! :\! PS,\! value\! :\! V\!\! >\!  PORTS2\! >
                  REST2 >
       REST) 
  =   
    portFixPoints(
       < O : CompositeActor | 
                ports : < P : OutPort | status : PS, value : V >  PORTS,
                innerActors : (O' ! P') ==> (\emph{parent} ! P ; EPIS)
                              < O' : Actor | ports : < P' : OutPort | > PORTS2 >
                              REST2 >
       REST)  \emph{if PS =/= unknown} .
\end{alltt}
\normalsize

An @owise@ equation is again used to end the fixed-point 
computation
when no equation adding new information about the ports can be applied.
However, to end the fixed-point computation of a (sub)system, the fixed-point
 computations of the subsystems of composite actors must have finished.
Therefore, this @owise@ equation should only be applied when 
there is no @portFixPoints@ operator in the @innerActors@ of the 
@CompositeActors@ in the system. Since @portFixPoints@ is declared as a \emph{partial}
 function, no object with an occurrence  @portFixPoint@ operator somewhere
in its inner actors (or in some subactor of an inner actor) will be a term of sort @Object@. 
That is, actors of sort @Object@ do not contain @portFixPoints@:

\small
\begin{alltt}
ceq portFixPoints(OBJECTS) = OBJECTS [owise] .
\end{alltt}
\normalsize

\paragraph{Modal Models.}
Most of the  semantics for modal models is borrowed from the semantics of  composite actors,
except for frozen actors, coupled ports,
and the evaluation order between the controller and refinements.
For modal models,
@postfire@  also
sets the @status@ attribute of the inner actors according to the current state of the controller
 to freeze all refinement actors except the refinement  of the current state:

\small
\begin{alltt}
eq postfire(
    <\! O\! :\! ModalModel\! |\! status\! :\! enabled, controller\! :\! CO, refinement\! :\! REFS,
                 \,    innerActors\! : CF\! >)
 = 
    <\! O\! :\! ModalModel\! |\! innerActors\! :\! (<\! CO\! :\! FSM-Actor\! |\! >
                                    setStateRefinement(STATE,\! REFS,\! ACTS))\! >\! 
   if <\! CO\! :\! FSM-Actor\! |\! currStatus\! :\! STATE\! >\!  ACTS  :=  postfire(CF) .
\end{alltt}
\normalsize

\noindent The function @setStateRefinement@ disables all refinements  except the
refinement of the current state.
%
%\small
%\begin{alltt}
%op setStateRefinement : Location RefinementSet Configuration -> Configuration .
%eq setStateRefinement(STATE, refine-state(STATE', O) REFS,
%                      < O : Actor | status : ACTSTAT > REST)
% = < O : Actor | status : if STATE == STATE' then \emph{enabled} else \emph{disabled} fi >
%   setStateRefinement(STATE, REFS, REST) .
%eq setStateRefinement(STATE, empty, REST) = REST .
%\end{alltt}
%\normalsize

If the controller depends on the result of @portFixpoints@ of  some refinement actors, then the result
must be transferred through some coupled input port of the controller actor. Hence 
the evaluation order between the controller and refinements is automatically treated in our representation.
The only part not yet covered  is to handle coupled input/output ports
in the controller FSM actor of a modal model.
In our representation, the coupled input/output ports have the same name,
and the value of the input port will be copied only if the coupled output port is \emph{absent}:

\small
\begin{alltt}
eq portFixPoints(
      < O : ModalModel | status : \emph{enabled},  controller : CO,
                         innerActors : 
                           < CO : FSM-Actor | status : \emph{enabled},
                                 ports : < P : InPort | status : \emph{present}, value : V\! > 
                                         < P : OutPort | status : \emph{absent}\! > PORTS >
                           REST2 >
      REST) 
 =   
   portFixPoints(
      < O : ModalModel | innerActors : \emph{portFixPoints(}
                          < CO : FSM-Actor | 
                              ports : <\! P\! :\! InPort\! |\! > 
                                      <\! P\! :\! OutPort\! |\! status\!\! :\! \emph{present}\!,\! value\!\! :\! V\!\! > PORTS\! >
                          REST2 >)
      REST) .
 \end{alltt}
\normalsize

\noindent The above equation can be only applied after the inner fixed-point computation
triggered by the controller FSM actor has been finished.
Therefore, an output port copies a value from its coupled input port only if
no value is generated at the output port when the controller is computed.

However, because of the above equation, the absent status of coupled output ports should not be
transferred to the parent until we can decide whether the associated coupled input port is absent or not.
For this reason we do not explicitly represent the connections between coupled output ports
of the controller and the output ports of the parent modal model.
Instead, we define the following equations to propagate the value of the coupled output ports:

\small
\begin{alltt}
eq portFixPoints(
     < O : ModalModel | status : enabled, controller : CO,
                        ports : < PI : OutPort | status : \emph{unknown} > PORTS,
                        innerActors : < CO : FSM-Actor | ports :
                                 < PI\! :\! OutPort\! |\! status\! :\! \emph{present}, value\! :\! V\! > PORTS'\! >
                                      ACTS >
     REST)
  =
   portFixPoints(
     < O\! :\! ModalModel\! |\! ports\! : <\! PI\! :\! OutPort\! |\! status\! :\! \emph{present}, value\! :\! V\! > PORTS >
     REST) .
\end{alltt}
\normalsize

\noindent The absent status of a coupled output port is propagated  only if the associated input port is also absent:

\small
\begin{alltt}
eq portFixPoints(
      < O : ModalModel | status : enabled, controller : CO,
                         ports : < PI : OutPort | status : \emph{unknown} > PORTS,
                         innerActors :
                            < CO : FSM-Actor |
                                 ports : <\! PI\! :\! InPort\! |\! status\! :\! \emph{absent}\! >
                                         <\! PI\! :\! OutPort\! |\! status\! :\! \emph{absent}\! > PORTS' >
                            ACTS >
      REST)
  =
   portFixPoints(
       < O : ModalModel | ports : < PI : OutPort | status : \emph{absent} > PORTS >  REST)\! .
\end{alltt}
\normalsize

\section{Specifying Temporal Logic Properties in Hierarchical Models}
\label{sec:temp-logic}
In Real-Time Maude, an LTL formula is constructed from a 
set of (possibly parametric) \emph{atomic state propositions} and the usual Boolean and
LTL operators. Having to define 
atomic state propositions makes the verification process nontrivial for
the Ptolemy user, since it 
requires some knowledge of the Real-Time Maude representation of
the Ptolemy model, as well as the ability to define  functions in Real-Time Maude. 
To free the user from this burden, we have predefined some generic 
atomic propositions, which extend the ones for flat Ptolemy models
in~\cite{icfem09}. 
For example, the property
\[ \mathit{actorId}  @ | @ \mathit{var_1}\: @=@ \:
\mathit{value_1}@,@ \ldots @,@\: \mathit{var_n}\: @=@\: \mathit{value_n}\] 
holds in a state if the value of  parameter $var_i$ of an actor equals $value_i$
for each $1 \leq i \leq n$, where
$\mathit{actorId}$ is the \emph{global actor identifier} of a given actor.
For FSM actors, the property
\[ \mathit{actorId}  \verb+ @ + \mathit{location} \] 
holds if and only if the FSM actor
with global name $\mathit{actorId}$ is in location (or ``local state'') $\mathit{location}$.

An LTL formula may contain multiple occurrences of the above atomic
propositions. To avoid having to unnecessarily write long global actor names
too many times, we can  simplify a formula for inner actors with an \emph{actor scope}, 
so that 
\[ \mathit{actorId}\;\: @:@\;\: \mathit{formula} \]
denotes that $\mathit{formula}$ should hold in the actor 
with the global identifier $\mathit{actorId}$. For example, the formula
$o_1 @.@\, o_2\;@:@\; @[]@\;( o_3 \;\verb+@+\; l_1\; @/\@\; o_4\, @.@\, o_5\; \verb+@+\; l_2)$ 
equals the formula
$@[]@\;(o_1 @.@\, o_2\,@.@\, o_3 \;\verb+@+\; l_1\; 
@/\@\;o_1\, @.@\, o_2\,@.@\, o_4\, @.@\, o_5\; \verb+@+\; l_2)$.

\section{Example: A Fault-Tolerant Hierarchical Traffic Light}
\label{sec:case-study}

This section shows how  a \emph{hierarchical} Ptolemy II DE model, that specifies
a  fault-tolerant traffic light system at a pedestrian crossing,
can be verified \emph{from within Ptolemy} using  Real-Time Maude. 
The Ptolemy II model is taken 
 from~\cite{BrooksFengLeevonHanxleden08_MultimodelingPreliminaryCaseStudy}.
The system consist of one car light and one pedestrian light.

\begin{figure}[h]
\begin{center}
\includegraphics[scale=0.88]{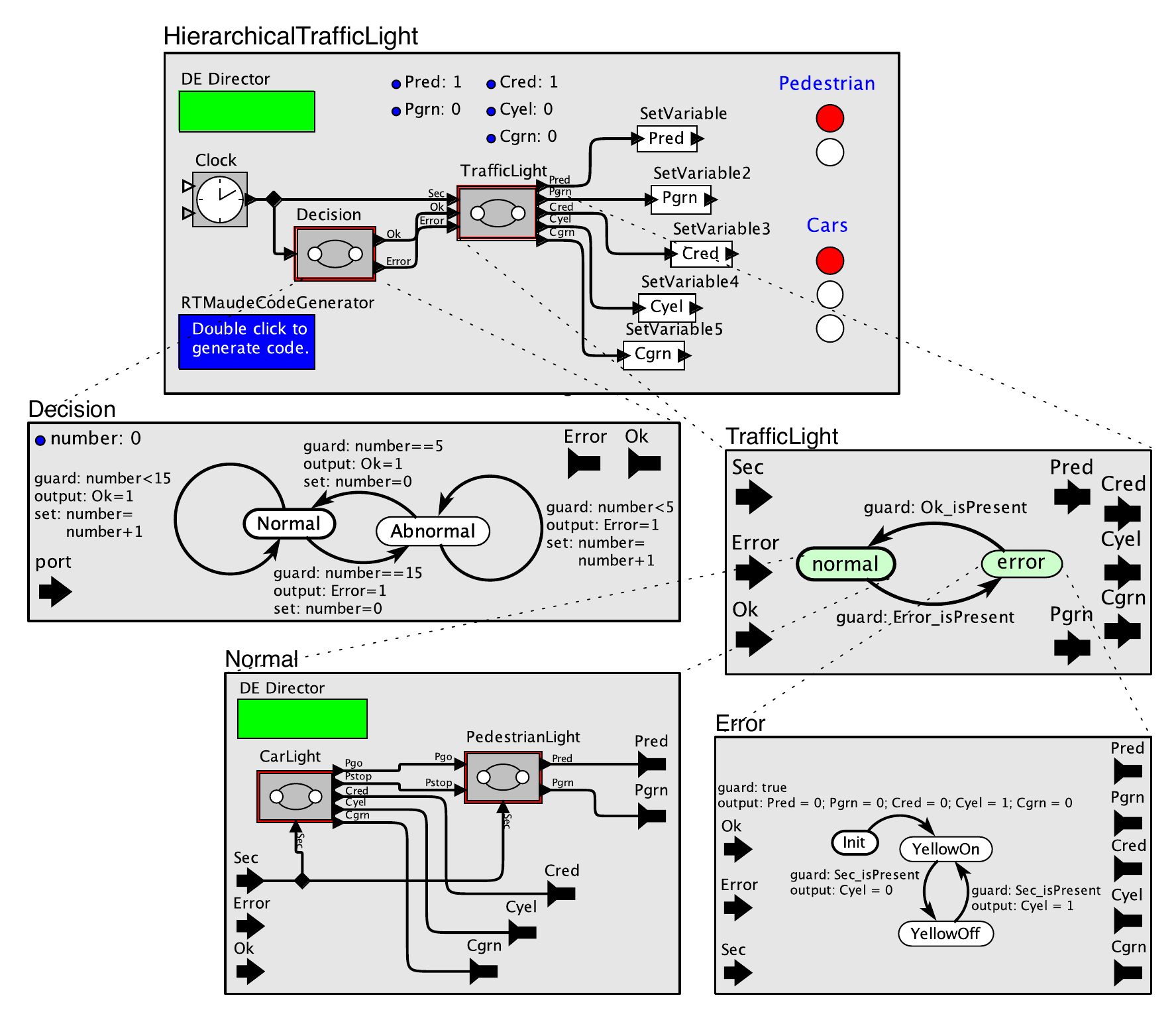}
\caption{A hierarchical fault-tolerant traffic light system. \label{fig:traffic1}}
\end{center}
\end{figure}

Figure~\ref{fig:traffic1} shows the system.
 The FSM
actor @Decision@ ``generates''  failures  and repairs by alternating between staying in
location @Normal@ for 15 time units and staying in location 
@Abnormal@ for 5 time units. Whenever the actor takes a transition
with target @Normal@, it sends a signal through its @Ok@ port, and
whenever it reaches, or stays in, location @Abnormal@, the actor sends
a signal through its @Error@ port.  @TrafficLight@ is a
modal model handling the two lights; 
whenever it is in @error@ mode and receives a signal through its @Ok@ port, 
the actor goes to @normal@ mode, and vice versa when it receives an @Error@
event in @normal@ mode. The FSM actor that refines the @error@ mode of @TrafficLight@ has
three states.
In this mode, all lights are turned off
(by sending a value 0 through the corresponding port), except for the yellow
light of the car light, which is blinking. 
The refinement of the @normal@ mode in @TrafficLight@ is the composite
actor that consists of the two FSM actors @CarLight@ and @PedestrianLight@,
that  define the behavior of the two lights during normal operations, and that have already been explained
in Section~\ref{sec:pt-example}. As before, 
 @Pred@, @Pgrn@, @Cred@,
@Cyel@, and @Cgrn@ are  variables that denote the current color(s) (if
any) of the lights.
Finally, the actor @Clock@ produces a signal every  time unit.

\begin{figure}[h]
\center
\includegraphics[scale=0.58]{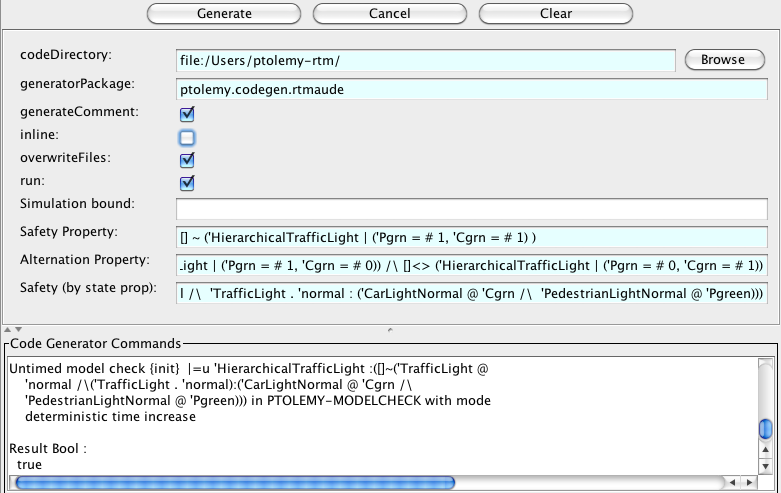}
\caption{The dialog box for the Ptolemy II verification\label{fig:dialog}}
\end{figure}

As explained in~\cite{icfem09}, we have used Ptolemy's  code generation infrastructure
to integrate \emph{both} the synthesis of a Real-Time Maude  model from a Ptolemy II
 model, \emph{and} the verification of the generated Real-Time Maude model, into Ptolemy II.
Figure~\ref{fig:dialog} shows the dialog box
that appears when the user double clicks on the
blue @RTMaudeCodeGenerator@ button in Ptolemy II DE models. 
This dialog box allows the user to write his/her model checking commands, and 
 then displays both the generated Real-Time Maude code and the results
 of the verification when the user presses the 'Generate' button. 

The main safety property is that  the car light and the pedestrian light never 
show green at the
same time. That is,  the variable @Pgrn@
and the variable @Cgrn@ should never be  1 at the same time, which, assuming that 
the  entire model is called @HierarchicalTrafficLight@, can be defined as the LTL formula

\small
\begin{alltt}
[] ~ ('HierarchicalTrafficLight | ('Pgrn = # 1, 'Cgrn = # 1) )
\end{alltt} 
\normalsize

We can also specify the safety of the traffic light controller.
If the traffic light
is in  @normal@ mode, the @CarLightNormal@ FSM actor should not be in state @Cgrn@ when 
the pedestrian light is in state @Pgreen@:

\small
\begin{alltt}           
'HierarchicalTrafficLight : (
  [] ('TrafficLight @ 'normal -> 
     ~ ('TrafficLight\! .\! 'normal : ('CarLight\! @\! 'Cgrn\, /\char92\!\!  'PedestrianLight\! @\! 'Pgreen))))
\end{alltt}
\normalsize

We can also check the liveness property that both pedestrian and cars
can cross  infinitely often. That is, it is infinitely
often the case the pedestrian light is green when the car light is \emph{not}
green, and it also  infinitely
often the case the car light is green when the pedestrian  light is
not green:

\small
\begin{alltt}
     []<> ('HierarchicalTrafficLight | ('Pgrn = # 1, 'Cgrn = # 0)) 
  /\char92 []<> ('HierarchicalTrafficLight | ('Pgrn = # 0, 'Cgrn = # 1))
\end{alltt}
\normalsize

\section{Related Work and Concluding Remarks}
\label{sec:related}

%% Patrick's work on SR to SMV: differences are: DE vs. SR; they don't
%% use the code gen helper framework; 
%% 
%% Maude semantics for modeling languages for embedded systems: Peter will do that.

A preliminary exploration of translations of \emph{synchronous
  reactive} (i.e., untimed) Ptolemy~II models into Kripke structures, 
that can be  analyzed by the NuSMV model checker, and of DE models into 
communicating timed automata is given
in~\cite{Cheng:08:AppliedVerification}. However, they 
require \emph{data abstraction} to map models into finitary 
automata, and they  flatten hierarchical models. 
On the other hand, as mentioned in the introduction, Real-Time
Maude has been used to define the semantics of several real-time languages, 
 but we are not aware
of any translation  of a synchronous and hierarchical real-time language into Maude or
Real-Time~Maude. The semantics of \emph{non-hierarchical} Ptolemy II DE models
is described in the recent paper~\cite{icfem09} that we extend to hierarchical models in this paper.

%% \section{Concluding Remarks}
\label{sec:concl}

We have shown how the Real-Time Maude formalization
of the semantics for flat Ptolemy II DE models has been extended
to the hierarchical case. Combining a fixed-point synchronous semantics with
hierarchical structure is not entirely trivial, as we explain in 
Section~\ref{sec:hierarchical-semantics}.  An additional  benefit of our work 
is the clarification  of the semantics of 
modal models, for which  we have also given  a composite-actor semantics in Ptolemy~II. 

We have  integrated Real-Time Maude code generation and model 
checking of hierarchical DE models into Ptolemy II, enabling
a model-engineering process for embedded systems that leverages the convenience of
Ptolemy~II DE modeling and simulation with the formal 
verification of Real-Time Maude. 

\small
\textbf{Acknowledgments.} 
This work is part of the Lockheed Martin Advanced Technology
Laboratories' NAOMI project. We thank
 the members of the NAOMI project
for enabling and encouraging this research;
 Christopher Brooks, Chihhong
Patrick Cheng, Thomas Huining Feng, Edward A.\ Lee, Man-Kit Leung, and Stavros Tripakis for discussions on
 Ptolemy~II; 
 Jos\'e Meseguer for  encouraging us to study the formal
semantics of Ptolemy in Real-Time Maude; and the anonymous reviewers
for  many helpful comments on an earlier version of this paper. 
We gratefully acknowledge financial support by
Lockheed Martin Corporation, through the NAOMI
project,  The Research Council of Norway, through the Rhytm project, and  NSF Grant CNS 08-34709.

\bibliographystyle{eptcs} 
\bibliography{bibl,ptolemy-bibl,Refs}

\end{document}